\documentclass[sigconf]{acmart}

\AtBeginDocument{%
  }

\setcopyright{acmlicensed}
\acmConference[EASE 2026]{The 30th International Conference on Evaluation and Assessment in Software Engineering}{Tue 9 - Fri 12 June 2026}{Glasgow, United Kingdom}

\copyrightyear{2026}
\acmYear{2026}
\setcopyright{cc}
\setcctype{by}
\acmConference[EASE '26]{International Conference on Evaluation and Assessment in Software Engineering}{June 09--12, 2026}{Glasgow, United Kingdom}
\acmBooktitle{International Conference on Evaluation and Assessment in Software Engineering (EASE '26), June 09--12, 2026, Glasgow, United Kingdom}
\acmDOI{10.1145/3816483.3816579}
\acmISBN{979-8-4007-2348-3/2026/06}

\begin{document}

\title{Assessing REST API Test Generation Strategies with Log Coverage}

\author{Nana Reinikainen}
\affiliation{%
  \department{Department of Computer Science}
  \institution{University of Helsinki}
  \city{Helsinki}
  \country{Finland}}
\email{nana.reinikainen@helsinki.fi}

\author{Mika M{\"a}ntyl{\"a}}
\affiliation{%
  \department{Department of Computer Science}
  \institution{University of Helsinki}
  \city{Helsinki}
  \country{Finland}}
\email{mika.mantyla@helsinki.fi}

\author{Yuqing Wang}
\affiliation{%
  \department{Department of Computer Science}
  \institution{University of Helsinki}
  \city{Helsinki}
  \country{Finland}}
\email{yuqing.wang@helsinki.fi}

\renewcommand{\shortauthors}{Reinikainen et al.}

\begin{abstract}
  Assessing the effectiveness of REST API tests in black-box settings can be challenging due to the lack of access to source code coverage metrics and polyglot tech stack. 
  We propose three metrics for capturing average, minimum, and maximum log coverage to handle the diverse test generation results and runtime behaviors over multiple runs.
  Using log coverage, we empirically evaluate three REST API test generation strategies, Evolutionary computing (EvoMaster v5.0.2), LLMs (Claude Opus 4.6 and GPT-5.2-Codex), and human-written Locust load tests, on Light-OAuth2 authorization microservice system. On average, Claude Opus 4.6 tests uncover 28.4\% more unique log templates than human-written tests, whereas EvoMaster and GPT-5.2-Codex find 26.1\% and 38.6\% fewer, respectively. Next, we analyze combined log coverage to assess complementarity between strategies. Combining human-written tests with Claude Opus 4.6 tests increases total observed log coverage by 78.4\% and 38.9\% in human-written and Claude tests respectively. When combining Locust tests with EvoMaster the same increases are 30.7\% and 76.9\% and when using GPT-5.2-Codex 26.1\% and 105.6\%. This means that the generation strategies exercise largely distinct runtime behaviors. Our future work includes extending our study to multiple systems.  
\end{abstract}



\maketitle

\section{Introduction}
The growing adoption of microservice architecture and polyglot tech stacks in modern web software development has increased the need for efficient and effective testing of REST APIs. For REST API testing, multiple tools exists \cite{MartinLopez2021,Atlidakis2019, schemathesis,Liu2022,Viglianisi2020,arcuri2019restful} and LLM-based test generation has been proposed \cite{kim2025llamaresttest, nooyens2025test}.

In black-box REST API testing, the test evaluation is typically based on metrics such as request counts, response status codes, and endpoint coverage. While these metrics reflect external API-level behavior, they provide limited insight into the diversity of runtime behaviors exercised by a test suite. Code coverage is a widely used metric in software testing to determine how extensively a system has been tested. However, computing code coverage usually requires source code or bytecode instrumentation, which is often impractical in microservice environments. Source code may not be accessible, instrumentation can introduce execution overhead, and deploying code coverage tools can expose engineering challenges \cite{LogCoCo}.

To address this limitation, we propose \emph{log coverage} as a coverage metric in black-box REST API testing. Logs are widely available and provide observable runtime information. We define log coverage as the number of distinct log templates, observed during test execution, where a log template represents a unique structured log message pattern produced by, for example, a log parser such as Drain \cite{Drain}. We consider each distinct log template as an indicator of differentiated runtime behavior, as log statements are typically associated with specific execution scenarios, system states, or processing branches, making them a reasonable proxy for behavioral diversity. Under this definition, log coverage quantifies the diversity of runtime behaviors exercised by a test suite. Importantly, log coverage does not require code instrumentation or system modification, making it particularly suitable for black-box environments.

The use of execution logs in software testing is rare, even though the concept was  proposed in the 1990s. The early work by \citet{logs-in-testing-90}  introduces state machines constructed from log events, which are used as test oracles and to support  testing. Prior work of \citet{LogCoCo} and \citet{log2cov} explores the relation between log coverage and code coverage, demonstrating that execution logs can be used as a proxy for code coverage.
\citet{reliability-test} proposes a method for automatically constructing Markov usage models and test cases based on logged user activity to test web service reliability.  \citet{load-tests} uses logs to automatically recover workloads for load testing. \citet{prioritization-log-div} introduce a log-based test case prioritization framework and implement seven techniques that combine log representations and three other prioritization strategies. This demonstrates that logs can serve as behavioral signals for guiding prioritization.
We found only two papers focusing on using logs in a microservice system context, both focusing on regression testing \cite{Corte2025Log,Chen2021Microservice}. Work by \citet{Chen2021Microservice} focused on the selection of regression tests, while \citet{Corte2025Log} trained a deep learning model that used execution logs to improve regression test suites. Finally, we found no papers where logs would have been used to evaluate test generation done by LLMs.

Given that LLM outputs in software testing require constant monitoring, \citet{harman2025harden,harman2025mutation} advocate for the development of “assured LLMs”, models whose outputs are accompanied by verifiable utility claims. We argue that software logs provide an excellent way to ensure the utility of LLM-created tests when metrics such as code coverage are not usable.

To assess the applicability of log coverage, we evaluate it with three representative REST API test generation strategies on a small industrial microservice system called Light-OAuth2, which implements the OAuth2 specification and consists of seven services. The strategies include: (i) EvoMaster, a search-based test generation tool using OpenAPI specifications, representing the state-of-the-art in automated coverage-oriented API testing; (ii) LLM test generation, an emerging approach that derives tests from semantic understanding of OpenAPI specifications leveraging latest LLMs (Claude Opus 4.6 and GPT-5.2-Codex) to generate test cases; and (iii) manually written Locust load tests, serving as a practitioner-designed baseline. For our evaluation, we define these research questions:

\begin{itemize}
    \item \textbf{RQ1}: How does log coverage differ across black-box API test generation strategies? 
    \item \textbf{RQ2}: Do different strategies reveal complementary runtime behaviors as measured by combined log coverage? 
\end{itemize}

To answer RQ1, we use log coverage as an estimate of behavioral exploration achieved by different test suites. By comparing the number of unique log templates discovered by each strategy, we assess their relative ability to exercise system behaviors. To answer RQ2, we use combined log coverage to assess whether different strategies reveal distinct runtime behaviors. The overlap in discovered unique log templates indicates shared behavioral exploration, where lower overlap suggests higher complementarity between strategies.

Our results indicate that LLM-generated tests by Claude Opus 4.6 discover nearly 30\% more unique log templates than the human-written Locust baseline, while EvoMaster-generated tests discover approximately 27\% fewer. Combining human-written tests with LLM-generated tests substantially increases overall coverage, demonstrating that the strategies exercise partially distinct runtime behaviors. Although our study is limited to a single system, it provides empirical evidence on the usefulness of log coverage for assessing and comparing black-box REST API test generation strategies.

This paper makes the following contributions:
\begin{enumerate}
    \item We propose log coverage, a log-based metric for approximating coverage in black-box API testing.
    \item We empirically compare three API test generation strategies using log coverage.
    \item We analyze the complementarity of these strategies through combined log coverage.
\end{enumerate}
\section{Methodology}
In this Section, we describe the methodology used for our empirical experiments, including  a description of the system under test, the API test generation strategies under evaluation, the experimental setup, and approaches used to assess complementary comparison.

\subsection{System Under Test}
The system under test (SUT) in this study is Light-OAuth2\footnote{\url{https://doc.networknt.com/service/oauth/}}, 
an open-source microservice implementation of the OAuth2 authentication and authorization specification developed by \emph{networknt}. The system consists of seven independent services that communicate via REST APIs and
support several database options. 
The available service API endpoints cover operations such as user login, user registration, service registration, client registration, issuing an access token, and public key certificate distribution.
In addition, the system includes the OpenAPI specifications for each of the services. These specifications define endpoint definitions, request schemas, and parameter constraints used for test generation.

\subsection{Test Generation Strategies}
\subsubsection{Locust}
Locust \footnote{\url{https://locust.io/}} is an open-source load testing framework for HTTP and other protocols. It provides an easy and flexible interface for defining basic create, read, update, and delete operations and more complex use cases of the SUT. In Locust, functions that contain API calls are defined as tasks. At the start of test execution, Locust spawns a virtual user population, where each repeatedly picks a random task to execute, sleeps for a predefined time, and picks a new task. Tasks can be weighted to control their execution frequency. Locust also supports scaling the number of concurrent users up to millions to simulate realistic load.

For this study, we use a manually written Locust test suite from the LO2 study targeting the Light-OAuth2 system \cite{LO2paper}. The test suite was designed based on the system's official API documentation and OpenAPI specifications, covering standard authorization flows and CRUD operations. It aims to achieve broad API coverage through multiple positive and negative test cases and enable log collection during test execution under both correct and erroneous API usage scenarios. As these tests were created without AI assistance, they required significant effort and a deep understanding of the system.

Due to differences in execution setup and concurrency characteristics, Locust tests serve as a reference baseline rather than a directly controlled experimental condition.

\subsubsection{EvoMaster}

EvoMaster \footnote{\url{https://github.com/WebFuzzing/EvoMaster}} is an open-source search-based API test generation tool \cite{arcuri2019restful}.  
We used the latest version of EvoMaster in black-box mode to automatically generate API test suites for SUT from its OpenAPI specification. In this black-box mode, EvoMaster applies evolutionary search techniques guided by observable API responses to explore API behaviors and increase endpoint coverage. The test generation process is formulated as a search problem, where the goal is to produce a test set that maximizes coverage while minimizing the number of test cases. The search process is guided by fitness functions that guide the exploration toward more comprehensive and efficient tests. Test cases are evolved from an initial population of randomly generated candidates and iteratively refined according to their fitness scores. The inputs of these test cases are derived from the SUT’s API specifications, which must be accurate and up to date to enable successful test generation. The search process terminates when a predefined stopping criterion is met, for example, execution time. The resulting test cases are self-contained and can be executed independently or composed into a full test suite.

\subsubsection{LLMs (Claude Opus 4.6, GPT-5.2-Codex)} We adopt an LLM-based strategy using two state-of-the-art models, Claude Opus 4.6 and GPT-5.2-Codex. We prompt these LLMs to produce executable API test cases according to the OpenAPI specifications of the SUT. This strategy leverages the generative capabilities of pretrained LLMs to synthesize test cases directly from specifications for a black-box setting. We access both models as out-of-the-box deployments via OpenRouter, without parameter tuning, model adaptation, or task-specific fine-tuning. We use the default configuration, including a fixed temperature of 1.0, which introduces controlled stochasticity into the generation process.

Although LLMs have recently attracted attention in software engineering research, relatively few studies \cite{kim2025llamaresttest, nooyens2025test}. have evaluated their use for API test generation directly from OpenAPI specifications in a black-box setting. Our approach therefore investigates LLM-based test generation as an emerging strategy that relies solely on publicly available API specifications, without access to source code or internal instrumentation.

\subsection{Experiment Setup}
To enable a fair comparison, we evaluate log coverage across three test generation strategies under consistent experimental conditions, controlling factors unrelated to the generation strategy itself.

\subsubsection{Test suite generation}
We generated test suites using the OpenAPI specification of the SUT as input for all automated approaches. We slightly modified the OpenAPI specification for the Client, Service, and User services by adding missing success status codes and adjusting parameters to match the actual service behavior.

\textit{For EvoMaster}, we executed it in black-box mode with a test generation budget of 70 seconds per service. We empirically determined that this time budget was sufficient to produce stable and representative test suites for Light-OAuth2. Increasing the time budget did not consistently yield better results.

\textit{For the LLM-based strategy}, we did not impose an explicit time limit. 
Instead, we recorded generation times and verified that they remained within a comparable range to those of EvoMaster (see Table~\ref{tab:test_generation_stats}). We applied the same test generation workflow to both LLMs to ensure a controlled comparison as follows. We first let LLMs generate an initial test suite based on the OpenAPI specification and execute it against the SUT. We allowed LLMs to refine failing tests once by feeding the test set along with the specification and the test execution report. The output formed the final test set. Finally, we manually performed a lightweight quality control to ensure, e.g., that tests use assertions in a meaningful way and not just increase coverage. Tests that did not pass the check were discarded. 

We designed the prompts for LLM-based generation using a structured six-component format comprising (1) role definition, (2) context, (3) instructions, (4) input data, (5) constraints, and (6) required output format \cite{ISTQB-CT-GenAI-v1.0}. 
The prompts were identical across LLMs and across runs and no model-specific adjustments were applied. A zero-shot strategy was adopted, meaning that no example test cases were provided in the prompt. To mitigate randomness effects and ensure robust results, we repeated test generation ten times for both EvoMaster and LLM-based approaches, following the recommended practices in \cite{randomness_eval}. For evaluation, all automatically generated test suites were executed against the SUT for 120 seconds under identical deployment, hardware, and logging configurations. \textit{For Locust},
we did not re-execute the test suite from the prior study \cite{LO2paper}, but randomly selected 10 test runs and corresponding log files as representatives of human-written test execution logs.

\subsubsection{Log collection and processing}
For each automated test generation strategy, we collect execution logs from running the final test suite against the SUT under a fixed logging configuration. For human-written tests, the logs were collected from the LO2 dataset.
Unique log templates were parsed from the logs and log coverage was computed based on their number, overlap, and distribution among services.

Before logs can be efficiently used for coverage estimation, they need to be loaded, masked, and transformed into a structured representation. This transformation process, log parsing \cite{Drain}, aims to group similar log entries that follow the same underlying pattern and constitute a log template by extracting constant parts of the log entry from dynamic runtime variables. For the purposes of this study, LogLead \cite{mantyla2024loglead} was used for loading and masking and Drain \cite{Drain} was selected as the log parser as it has been shown to produce accurate log templates with low parsing time latency \cite{parser-survey}. Custom LO2 specific masking was added to LogLead masks by empirically analyzing parsed Drain templates.

\subsection{Metrics}

\subsubsection{Average Log Coverage (AvgLC)}

To estimate the typical behavioral profile of a test generation strategy, we computed what we refer to as Average Log Coverage (AvgLC) in two steps. First, we compute the average number of unique log templates discovered across 10 independent runs. But, as we need a set, not just a number, we continue in the second step. There  we sort log templates from all runs (by each strategy) with decreasing frequency and select the average count of top templates. 

This frequency-based aggregation represents the average system behavior exercised by the test suites. Because AvgLC favors the most frequently appearing log templates across all runs, it reduces noise caused by infrequently appearing log templates. However, discarding rare outlier log templates may underestimate the exploration power of a test generation strategy.

\subsubsection{Minimum Log Coverage (MinLC)}
To estimate the lower bound of consistently reproducible log coverage of each test generation strategy, we computed Minimum Log Coverage (MinLC) as the intersection of unique log templates across all 10 independent runs. Only those log templates that appear in every individual run are included. MinLC captures deterministic and repeating runtime behavior while filtering out all rare events and execution paths that occur only sporadically. As a result, MinLC provides a conservative estimate of behavioral coverage and reflects the minimum behavior a strategy reliably exercises.

However, because MinLC excludes templates that do not appear in all runs, it may underestimate the true behavioral exploration capability of a test generation strategy even more than AvgLC.

\subsubsection{Maximum Log Coverage (MaxLC)}
To estimate the upper bound of behavioral exploration capability of each test generation strategy, we computed Maximum Log Coverage (MaxLC) as the full union of unique log templates across all 10 independent runs. For each strategy, all unique log templates discovered in individual runs are combined into a single aggregated set, representing the total observable behavior exercised across repetitions. Unlike AvgLC and MinLC, MaxLC captures every distinct runtime behavior identified during experimentation, including rare events and edge cases that may appear only in a subset of runs.

However, MaxLC may be inflated by infrequently occurring templates and is therefore more sensitive to randomness. As a result, it does not represent typical performance.

\subsubsection{Costs}
Here we outline test development effort and cost to the best of our knowledge. These allow us to understand the costs of different test generation strategies and, therefore, provide information for making decisions between test generation strategies. All human effort numbers are estimates by the test authors, given after the task was completed.

Locust test \cite{LO2paper} were developed during 30 working days with roughly 5 hours per day, resulting in a human effort of 150 hours. Computing costs for Locust are negligible as it runs on normal hardware. No LLMs were used to develop Locust tests. 

Using EvoMaster involves relatively low human effort after the correct configurations were found. The setup mainly consists of downloading the EvoMaster execution file, configuring the tool, deploying the SUT, and executing the generation process, after which a complete test suite is automatically produced. EvoMaster runs locally, meaning that computing costs are limited to local hardware resources and execution time. We estimate that the total human effort required throughout the setup, configuration, and execution process is approximately 4 hours.

LLM-based approaches required human effort in designing an effective prompt and performing quick manual quality checks for generated tests. Test generation was performed via external API calls through OpenRouter, which introduce usage-based pricing. The estimated cost for 10 runs is approximately \$13 for Claude Opus 4.6 and \$3 for GPT-5.2-Codex. We estimate the total human effort to be approximately 10 hours for prompt development (shared with models), plus 2 hours per model, including quality checks.

\section{Results}
\label{sec:res_rq1}
\subsection{RQ1: Log coverage across testing strategies}

\begin{table*}[t]
\centering
\caption{Test Statistics. Data is presented as AvgLC (MinLC--MaxLC).}
\begin{tabular}{llllll}
\hline
Method & Generation Time (s) & \# Tests & \# Unique Log Templates & Pass Ratio (\%) & Lines of Test Code\\
\hline
Locust          & NA              & 79 (79--79)    & 88 (84--92)    & NA                    & 1896 (1896--1896) \\
EvoMaster       & 707 (660--767)  & 54 (52--57)    & 65 (62--69)    & 88.14 (87.04--90.57)  & 1149 (1116--1206)\\
Claude Opus 4.6 & 447 (417--492)  & 177 (170--185) & 113 (97--120)  & 96.49 (93.02--99.43)  & 1851 (1717--1947) \\
GPT-5.2-Codex   & 743 (547--988)  & 32 (27--36)    & 54 (49--60)    & 94.64 (69.7--100)    & 385 (314--479)\\ 
\hline
\end{tabular}
\label{tab:test_generation_stats}
\end{table*}

\begin{table}[h]
\centering
\caption{Average Unique Log Templates Per Service}
\label{tab:average_unique_templates_rotated}
\begin{tabular}{lrrrr}
\toprule
Metric & Locust & EvoMaster & Claude 4.6 & GPT-5.2\\
\midrule
Client      & 31.8 & 23.0 & 36.9 & 16.5 \\
Code        & 44.0 & 28.0 & 34.4 & 24.1 \\
Key         & 4.8  & 17.3 & 23.0 & 17.2 \\
Ref.-token  & 19.7 & 16.3 & 24.6 & 15.7 \\
Service     & 25.8 & 23.9 & 32.0 & 15.3 \\
Token       & 28.2 & 22.2 & 48.9 & 18.4 \\
User        & 21.5 & 22.8 & 24.1 & 13.0 \\
\bottomrule
\end{tabular}
\end{table}

Table~\ref{tab:test_generation_stats} summarizes test generation characteristics. 

\textbf{Overall Log Coverage}. Claude Opus 4.6 achieved the highest log coverage with 113 unique log templates on average, followed by Locust (88), EvoMaster (65), and GPT-5.2-Codex (53). Compared to Locust tests, Claude Opus 4.6 improves log coverage by nearly 30\% and achieves more than twice the coverage of GPT-5.2-Codex.

\textbf{Log Coverage per Service.} As reported in Table~\ref{tab:average_unique_templates_rotated}, log coverage exhibits clear service-dependent differences across strategies. Claude Opus 4.6 tests achieve the highest log coverage in six out of seven evaluated services, being particularly strong in Key and Token services. The only exception is the Code service, where Locust outperforms all other strategies. EvoMaster generally remains below both Claude Opus 4.6 and Locust baseline in most services, while GPT-5.2-Codex achieves the lowest overall coverage. 

These service-level patterns help explain the coverage differences observed earlier. In particular, the Code service represents a notable exception to the overall trend. It implements complex authorization workflows that cannot be accurately defined in API specifications. This projects directly to a lower number of discovered unique log templates in the LLM and EvoMaster tests. In contrast, Locust executes predefined workflow sequences and therefore achieves higher coverage in the Code service. However, in other services, Locust primarily exercises typical execution paths. Claude Opus 4.6, by comparison, explores more diverse input combinations, edge cases, and less frequent behaviors, which is reflected in its higher log coverage across most services.  
    
\textbf{Factors affecting log coverage.} \textit{Test suite size.} The higher log coverage achieved by Claude Opus 4.6 is partially attributable to the larger size of the test suite. Although Claude Opus 4.6 generates substantially more tests on average, the relative increase in discovered unique log templates is not proportional. When approximating the number of unique log templates discovered per test, Claude Opus 4.6 achieves $113/177 \approx 0.64$, compared to EvoMaster ($65/54 \approx 1.20$), GPT-5.2-Codex ($53/32 \approx 1.66$), and Locust ($88/79 \approx 1.11$). This trend is also reflected in the total lines of generated test code: Claude Opus 4.6 produces 1851 lines on average, substantially more than GPT-5.2-Codex (385 lines) and EvoMaster (1149 lines), and comparable to Locust (1896 lines).

\textit{Test pass ratio}. Although log coverage is computed from all executed tests regardless of their verdict, pass ratio may affect coverage, as tests that terminate prematurely may exercise fewer execution paths. In our results, Claude Opus 4.6 achieves both high average pass ratio (96.49\%) and high log coverage (113 templates). However, GPT-5.2-Codex exhibits substantial variability in pass ratio across runs (69.7\%–100\%), likely due to differences in the generated test sets, while maintaining the lowest average log coverage (53 templates).

\textit{Test generation variability} among the strategies also affects the log coverage achieved by each test generation strategy. LLM-based tests show greater variability in both test suite size and pass ratio across runs compared to EvoMaster. This variability directly impacts the number of unique log templates discovered: Claude Opus 4.6 consistently finds more templates than Locust, although the results vary greatly between runs, while GPT-5.2-Codex occasionally reaches coverage comparable to EvoMaster’s minimum, but generally produces fewer templates.

\textit{Test generation time} does not exhibit a positive relationship with log coverage in our results. Claude Opus 4.6 achieves the highest coverage (113 templates) despite requiring less generation time (447s) than EvoMaster (707s) and GPT-5.2-Codex (743s), both of which produce lower coverage.

\subsection{RQ2: Complementary gains in log coverage}

\begin{table*}[t]
\centering
\caption{Combined Log Coverage Comparison showing AvgLC (MinLC--MaxLC). Note that percentages for MaxLC and MinLC are not the same as min and max in statistics.}

\begin{tabular}{llll}
\toprule
Metric & Locust vs. EvoMaster & Locust vs. Claude Opus 4.6 & Locust vs. GPT-5.2-Codex \\
\midrule
Unique Log Templates (Locust)    & 88 (68--160)         & 88 (68--160)               & 88 (68--160)             \\
Unique Log Templates (Other)     & 65 (34--188)         & 113 (48--358)              & 54 (17--133)             \\
Intersection of Templates        & 38 (24--52)          & 44 (32--59)                & 31 (12--49)              \\
Union of Templates               & 115 (78--296)        & 157 (84--459)              & 111 (73--244)            \\
Jaccard Similarity (\%)          & 33.04 (30.77--17.57) & 28.03 (38.10--12.85)       & 27.93 (16.44--20.08)     \\
Gain over Locust (\%)            & 30.68 (14.71--85.00) & 78.41 (23.53--186.88)      & 26.14 (7.35--52.50)      \\
Gain over Other (\%)             & 76.92 (129.41--57.45)& 38.94 (75.00--28.21)       & 105.56 (329.41--83.46)   \\
\bottomrule
\end{tabular}
\label{tab:combined_behavior_comparison}
\end{table*}

Table~\ref{tab:combined_behavior_comparison} presents a comparison of human-written Locust tests with other test generation strategies.

\subsubsection{Jaccard Similarity}
Table~\ref{tab:combined_behavior_comparison} shows the unique log templates identified in both the Locust and the respective (Other) test generation approach. After that, the table outlines the intersection and union of these log template sets. Then Jaccard similarity is shown, calculated as the intersection divided by the union.

Jaccard similarity in the average case is 33\%, 28\% 28\%, when comparing Locust against Evomaster, Claude Opus 4.6 and GPT-5.2-Codex respectively. The observed range spans from a minimum of 13\% to a maximum of 38\% when considering MinLC and MaxLC. Overall, Jaccard similarity indicates that the generated test sets exhibit substantial dissimilarity when compared to the human-written Locust set.

\subsubsection{Gain when combining the approaches}

The "Gain over Locust" row indicates the percentage increase in unique templates obtained by supplementing the human-written Locust tests with each respective test generation method. This value is computed by dividing the union of templates by the unique templates identified by Locust alone.

Claude Opus 4.6 demonstrates the highest gain, contributing an additional 78\% of unique log templates. This is consistent with its superior log coverage, as shown in Section \ref{sec:res_rq1}. In contrast, EvoMaster and GPT-5.2-Codex yield more modest gains of 31\% and 26\%, respectively. 

When examining the minLC and maxLC measures, the "Gain over Locust" metric exhibits substantial variability across setups. When adding Claude Opus 4.6 test to Locust, the gain ranges from a minimum of 24\% to a maximum of 187\%. Even EvoMaster, considered the most stable approach, shows a spread of 15\% to 85\%, while GPT-5.2-Codex ranges from 7\% to 52.5\%. 

Reversing the perspective, i.e., assessing the benefit of supplementing each test generation strategy with the Locust template set, reveals that even Claude Opus 4.6, despite having the highest baseline coverage, still gains an average of 39\% additional unique templates. For EvoMaster, integrating Locust templates yields a 77\% gain on average, while GPT-5.2-Codex experiences the most significant augmentation, with a 106\% increase in unique log templates.

An analysis of minLC and maxLC measures further reveals substantial variation. Claude Opus 4.6 exhibits a gain ranging from 28\% (maxLC) to over 75\% (minLC) when supplemented with Locust templates. The spread is even more pronounced for GPT-5.2-Codex, with gains spanning from 80\% to 330\%, while EvoMaster ranges between 57\% and 129\%. 

\subsubsection{Summary}
The Jaccard similarity and the "Gain over" analyses collectively demonstrate that the test generation strategies are largely complementary. While Claude Opus 4.6 achieves the highest baseline coverage, as established in Section~\ref{sec:res_rq1}, it still has limitations. Other approaches are behind Locust, but can still provide meaningful log template coverage increases to it. This complementarity suggests that combining multiple approaches yields more robust test suites.

\section{Limitation and Future work}
This study has limitations related mainly to logging quality and the selected environment and setup. Future work directions are outlined to address the limitations.

The experiments were performed only on Light-OAuth2, whose architecture and logging practices may differ from other REST-based microservice systems. Experimenting with a single system limits the external validity of the proposed method. Future work includes extending this study to multiple systems. 

Drain was the only log parser used for log template extraction. Different parsers may produce different template sets, which could alter the coverage values.

Logging distribution, density, and consistency also influence observed log coverage. Systems with sparse, overly dense, or inconsistent logging may yield misleading coverage values. This affects the validity and reliability of log coverage.

Absolute log coverage based on actual logging statements was not computed. While this would have provided a clearer comparison of test strategies, the study focused on a black-box approach.

The prompting approach may have influenced the quality of LLM-generated tests. Excessive prompt engineering was not performed. However, the same prompts and setup were used for both LLMs. Difference between Claude and GPT may stem from Claude's stronger specification-based test generation abilities.

We are also planning other ways of assessing test strategies output such a qualitative analysis of the (generated) tests and asserts.

To facilitate the reproducibility of this study, we provide the replication package in Zenodo: https://doi.org/10.5281/zenodo.19944008
\section{Conclusion}
In this study, we evaluated different REST API test generation strategies in black-box settings on the LO2 microservice system.  
As our baseline, we utilized human-written Locust tests and compared it with EvoMaster, a traditional search-based API test generation tool, as well as two leading LLMs available at the time: Claude Opus 4.6 and GPT-5.2-Codex.

Our findings indicate that Claude Opus 4.6 outperforms all other strategies in terms of log coverage. Human-written tests ranked the second, followed by EvoMaster and GPT-5.2-Codex. Notably, relying on a single test generation strategy resulted in limited log coverage. When combining the two best-performing test generation strategies, Claude Opus 4.6 and  human-written tests, we observed an additional 39\%  increase in coverage beyond what Claude Opus 4.6 achieved alone. This highlights the complementary nature of these strategies in maximizing log coverage.

\begin{acks}
This work is supported by the Research
Council of Finland (grant id: 359861, the MuFAno project). This work is also funded by the EuroHPC Joint Undertaking and its members including top-up funding by the Ministry of Education and Culture of Finland.
\end{acks}

\bibliographystyle{ACM-Reference-Format}
\bibliography{references}

\end{document}